\documentclass[12pt]{article}
\usepackage{graphicx}
\usepackage{units}
\usepackage{html}
\usepackage{units}
\usepackage{amssymb}
\usepackage{amsmath}
\usepackage{epsf,epsfig}

\newcommand{\daikonplus}{{\tt v3.5.5}}

\newcommand{\ngen}{{\tt NEUGEN3}}

\newcommand{\inuke}{{\tt INTRANUKE}}
\newcommand{\carrot}{{\tt carrot}}
\newcommand{\daikon}{{\tt daikon}}

\title{Hadronic Shower Energy Scale Uncertainty in the MINOS Experiment}
\author{Steve Dytman \\ \it{University of Pittsburgh} 
\and
Hugh Gallagher \\ \it{Tufts University}  \and
Michael Kordosky \\ \it{College of William and Mary}} 

\begin{document}
\maketitle

\begin{abstract}
In this paper we determine the model uncertainty in the calorimetric response of the MINOS detector 
to hadronic showers produced by neutrino interactions. 
\end{abstract}

\section{Introduction}
The uncertainty in the incoming neutrino energy plays an important role in long-baseline 
neutrino oscillation experiments, in particular the MINOS experiment \cite{Harris:2004iq,Kordosky:2006gt}.  
The determination of  $| \Delta m_{23}^2 |$ from a reconstructed charged current energy spectrum involves
model dependencies, even in the presence of a near detector.  One reason is that in order to determine
the neutrino energy for a given event, one needs a model for how much energy goes `missing' in the
nuclear environment.  This missing energy is strongly detector dependent and includes nuclear binding,
charged pion rest masses, and intranuclear absorption of produced hadrons.   The absorption of pions
can dramatically change the detector response since the pion energy is typically redistributed to
a sizable number of nucleons, which are often below the detection threshold.   In this note we 
will survey the components of a particular neutrino interaction model that contribute 
to this uncertainty and tabulate contributions to the hadronic shower energy scale
uncertainty.  The model used for this study is version \daikonplus{} of \ngen{} \cite{Gallagher:2002sf}, 
the event generator version used for the production of the 2008 round of Monte Carlo simulations by the MINOS 
experiment.    Details of the MINOS analysis can be found elsewhere \cite{Adamson:2007gu}.     

\section{The NEUGEN3 Neutrino Event Generator}
NEUGEN3 is a widely-used neutrino event generator that produces complete final states for neutrino-nucleus 
interactions for energies from 100 MeV to 100 GeV.  It incorporates a Fermi Gas Model as the basic 
nuclear model with some modifications for nucleon-nucleon correlations \cite{Bodek:1981ar}.  The cross section
model includes quasi-elastic interactions \cite{LlewellynSmith:1972zm}, 
resonance production \cite{Rein:1981wg}, 
coherent neutrino-nucleus scattering \cite{Rein:1983pf}, 
and non-resonant inelastic scattering.  The 
model for non-resonant inelastic scattering is a modified DIS model which can describe electron scattering 
structure function data down to very low $Q^2$ and was designed for use by neutrino experiments in the 
few-GeV energy range \cite{Bodek:2004pc}.  Although the events from this model often do not fall in the 
canonical DIS regime of Q$^2>$1 GeV$^2$ and W$>$2 GeV/c$^2$, we will nonetheless in this document refer 
to the class of events generated by this model as ``DIS events''. 
  The most important model aspects for the hadronic shower scale
uncertainty are the hadronization model, which determines the set of particles produced from a particular 
DIS event, the formation zone, and the intranuclear rescattering model, which determines how this set of hadrons is altered
as altered by final-state interactions (fsi) as they exit the target nucleus.    We will describe the 
hadronization, formation zone, and intranuclear rescattering models in somwehat more detail. 

\subsection{Hadronization Model}  
NEUGEN uses the AGKY model to describe the hadronization process from events 
produced by DIS interactions \cite{Yang:2007zzt}.   This model has two components.  The low invariant mass component
is called the modified-KNO model and is largely based on empirical parametrizations 
of hadronic system features like Koba-Nielsen-Olesen scaling
\cite{Gallagher:2002sf,Koba:1972ng}.  The high invariant mass component is JETSET \cite{Sjostrand:1993yb}.  
The modified-KNO model is used for all DIS events below W$=$2.3 GeV/c$^2$ and JETSET is used for 
all DIS events above W$=$3.0 GeV/c$^2$.  A linear transition is made between the two models for events with 
2.3 GeV/c$^2<$W$<$3.0 GeV/c$^2$. 

\subsection{Formation Zone}
Within the past few years results on the production of hadrons in a nuclear environment have
been released by HERMES \cite{Akopov:2007zz},
CLAS \cite{Hafidi:2006ig}, and JLab Hall C \cite{Clasie:2007gqa}.   It is now well established that 
it takes the hadrons time to `form' in the  nuclear medium and that during this time they interact
with a reduced cross section.   The modeling of this effect in \inuke{} is based on a study from 
the SKAT neutrino experiment \cite{Baranov:1984rv}.  We utilize a very simple model where it takes
the hadron a time $\tau_0$ to form during which its interaction cross section in the nucleus
is set to zero.  The value of $\tau_0$ is based on \cite{Baranov:1984rv} and is taken to be 
0.342 fm/c.     

\subsection{Intranuclear Rescattering}
Like all neutrino event generator packages, the MINOS model for intranuclear rescattering of produced
hadrons uses a semiclassical Intranuclear Cascade (INC) model.  This subpackage of \ngen{} is called 
\inuke{}  and has undergone numerous revisions and updates since it was originally written \cite{Merenyi:Thesis}.  
The simulation is anchored to a large body of hadron-nucleus scattering data and has been compared with the 
results of neutrino bubble chamber scattering on neon \cite{Merenyi:1992gf}.     

The intranuclear cascade simulation proceeds in two steps. In the first step each produced pion and nucleon 
is stepped through the nucleus to determine if it interacts.  The calculation at each point (r) uses the 
mean free path based on the local density:
\begin{equation}
\lambda(E,r)=\frac{1}{\rho(r)\sigma_{hN}(E)}
\end{equation}
where $\rho$ is the nuclear matter density and $\sigma_N$ in the total hadron-nucleon 
cross section ($\sigma_{tot}$).  The density distribution for
 nuclei heavier than oxygen are modeled with a Woods-Saxon
distribution, others are modeled with a Gaussian distribution.  

A fundamental problem exists in that we are trying to use 
a semiclassical model to describe a quantum
mechanical phenomenon.  The primary advantage is that the highly inelastic processes
such as pion absorption and inelastic scattering can be matched to data 
while quantum mechanical treatments are much more difficult
or impossible.  The primary disadvantage is that elastic scattering, a large contributor to 
the total scattering cross section at low energy, is primarily
a wave process.  
The result is that a  naive semiclassical 
model fails to correctly describe the measured hadron-nucleus cross sections at low 
hadron energy.  In \inuke{} we account for this by increasing the nuclear size in the 
transport calculation by a fraction of the hadron de Broglie wavelegnth.  We take this 
fraction to be 0.5 by comparing with $\pi+^{56}\mathrm{Fe}$ total cross section data. 

In \inuke{} each neutrino-produced hadron is allowed to reinteract in the nucleus only once.   
Once an interaction has taken place, a list of 
particles in the ``final state'' is determined and these particles are placed outside 
the nucleus.  The choice of final state is derived from data on hadron interactions with iron. 
Since iron is the principal nucleus in the MINOS detector this approach will, by design, give
the correct distribution of final states in iron at the expense of lacking the full generality 
of a complete INC model where multiple reinteractions are allowed to occur within the nucleus. 
The reaction types that are considered include charge exchange, elastic scattering, 
inelastic scattering, secondary pion production, and absorption and are implemented in the 
code essentially as energy-dependent branching ratios. 

\section{Methodology}

In this analysis we have determined the contribution to the hadronic shower energy scale uncertainty from 15 sources. 
The general methodology can be described as follows:
\begin{enumerate}
\item 
{\bf Identify Sources of Uncertainty:} 
The first task was to identify sources of uncertainty.    We can view the generator as a set of 
physics models which are tuned to some external data.    These models are developed in a certain theoretical
framework and may involve significant assumptions or approximations.  Likewise the data which is used for model tuning
may possess substantial measurement uncertainty.    
We can broadly categorize these classes of uncertainties into {\em external data}
uncertainties and {\em model} uncertainties.  
\item 
{\bf Quantify uncertainty:}
Once we have enumerated all of the individual sources of uncertainty we have to 
quantify these uncertainties.   For instance, what is the uncertainty on the pion absorption cross section itself?  For 
external data uncertainties these {\em input} uncertainties are fairly straightforward to evaluate for one familiar with the 
external data in question.  
\item 
{\bf Evaluate output uncertainty:}
All estimates presented here were determined by comparing the results of 
4-vector simulations combined with a parametrized detector response 
model.   The detector response model is a parametrization of the output of full GEANT3 simulations of the response of the 
MINOS detector to various particles over a range of energies.  
In each case 500k events were generated in the NuMI low energy beam.  A reference sample of events was generated 
using \ngen{} \daikonplus{}.    We then compute the change in the estimated response for a particular model change 
in bins of true shower energy.  
The shift in each bin is calculated as (shifted response - default response)/(default response).    
\end{enumerate}   

\section{Uncertainty due to final state interactions}

\subsection{External Data}
\begin{table}[t]
\centering
\begin{tabular}{|l|c|}
\hline
\multicolumn{2}{|c|}{branching ratios}\\
\hline
parameter & $1\sigma$ uncertainty (\%)\\
\hline
$\pi$ charge-exchange & 50 \\
$\pi$ elastic & 10 \\
$\pi$ inelastic & 40 \\
$\pi$ absorption & 30 \\
$\pi$ secondary $\pi$ production & 20 \\
N absorption & 20 \\
N secondary $\pi$ production & 20 \\
N elastic & 30 \\
\hline
\multicolumn{2}{|c|}{cross-sections}\\
\hline
parameter & $1\sigma$ uncertainty (\%)\\
\hline
$\pi$ total cross-section & 10 \\
N total cross-section & 15 \\
\hline
\end{tabular}
    \caption{  Uncertainties on intranuclear rescattering processes.  
The N elastic and total cross-section terms are 100\% correlated. \label{tab:inuke_unc}}
\end{table}

For this part of the study we have evaluated the effect of the ten sources of uncertainty listed in Tab.~\ref{tab:inuke_unc}. 
For this study we shifted each of these inputs by $+1\sigma$.    In the case of specific reaction 
cross sections, the other cross sections are scaled down in their original proportions so that the total scattering
cross section is unchanged.   In each case the stated uncertainty refers to the magnitude of the relevant branching ratio 
or cross-section. 
The underlying cross-sections and branching ratios are energy dependent as are their uncertainties.  
The values adopted here correspond to the maximum value of the energy dependent uncertainties and are therefore overestimates.

\begin{figure}
  \includegraphics[width=\columnwidth]{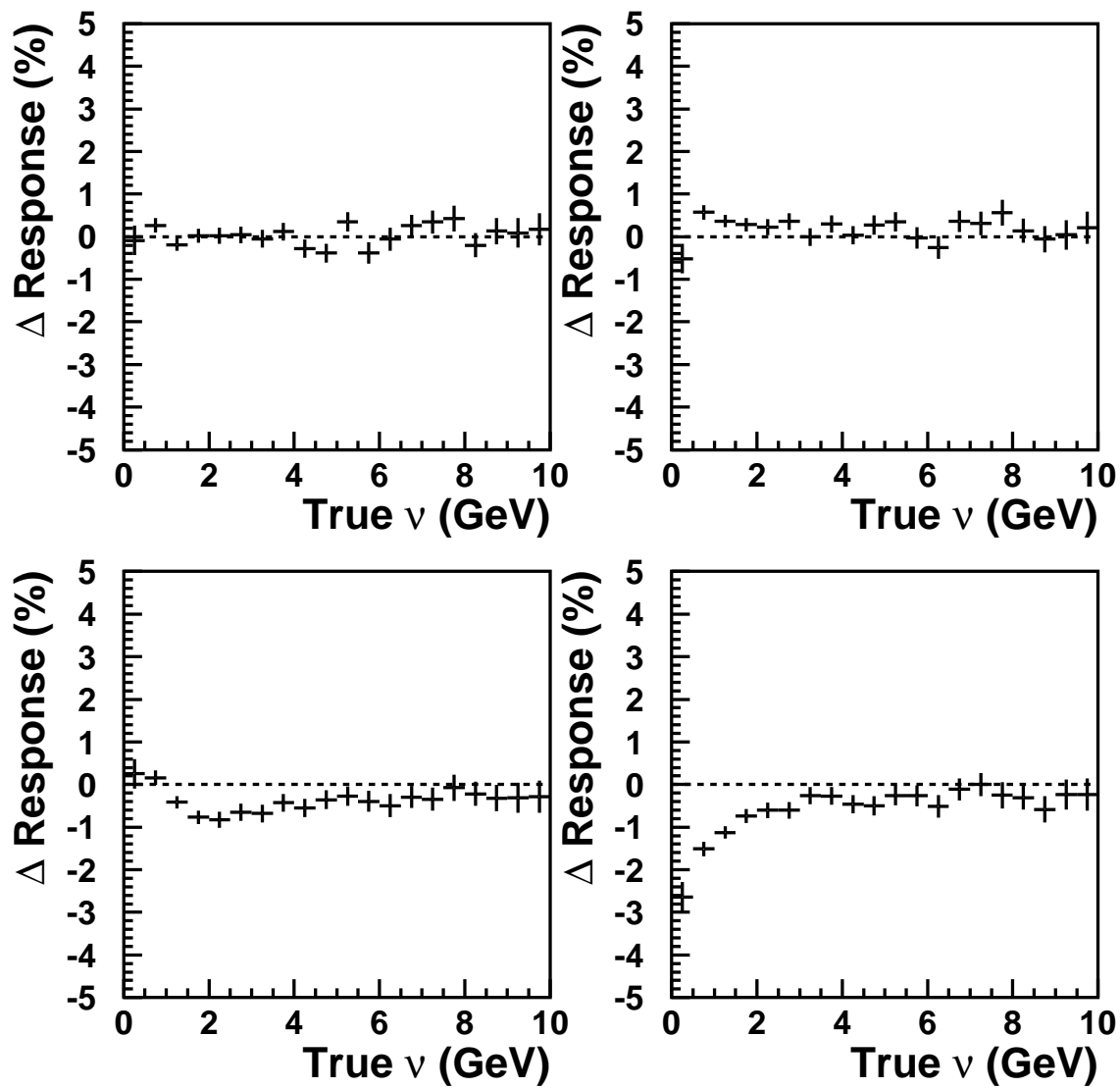}
  \caption{\label{fig:results2-5}The figures correspond to $+1\sigma$ shifts in $\pi$ charge exchange (top left), $\pi$ elastic (top right), $\pi$ inelastic (lower left) and $\pi$ absorption (lower right).}
\end{figure}

\begin{figure}
  \includegraphics[width=\columnwidth]{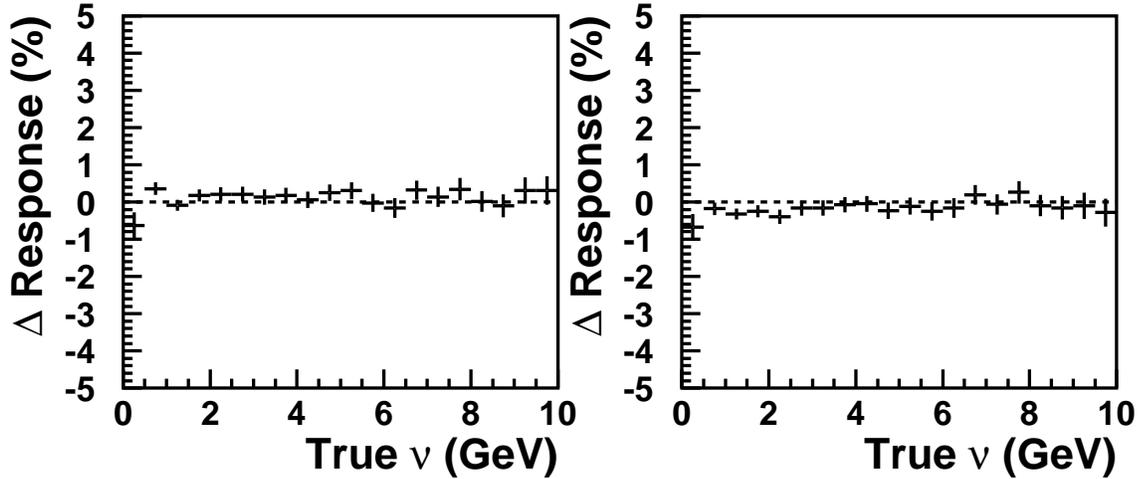}
  \caption{\label{fig:results6-7}The results of $+1\sigma$ shifts to pion secondary pion production (left) and 
the pion scattering cross section (right).}
\end{figure}

\begin{figure}
  \includegraphics[width=\columnwidth]{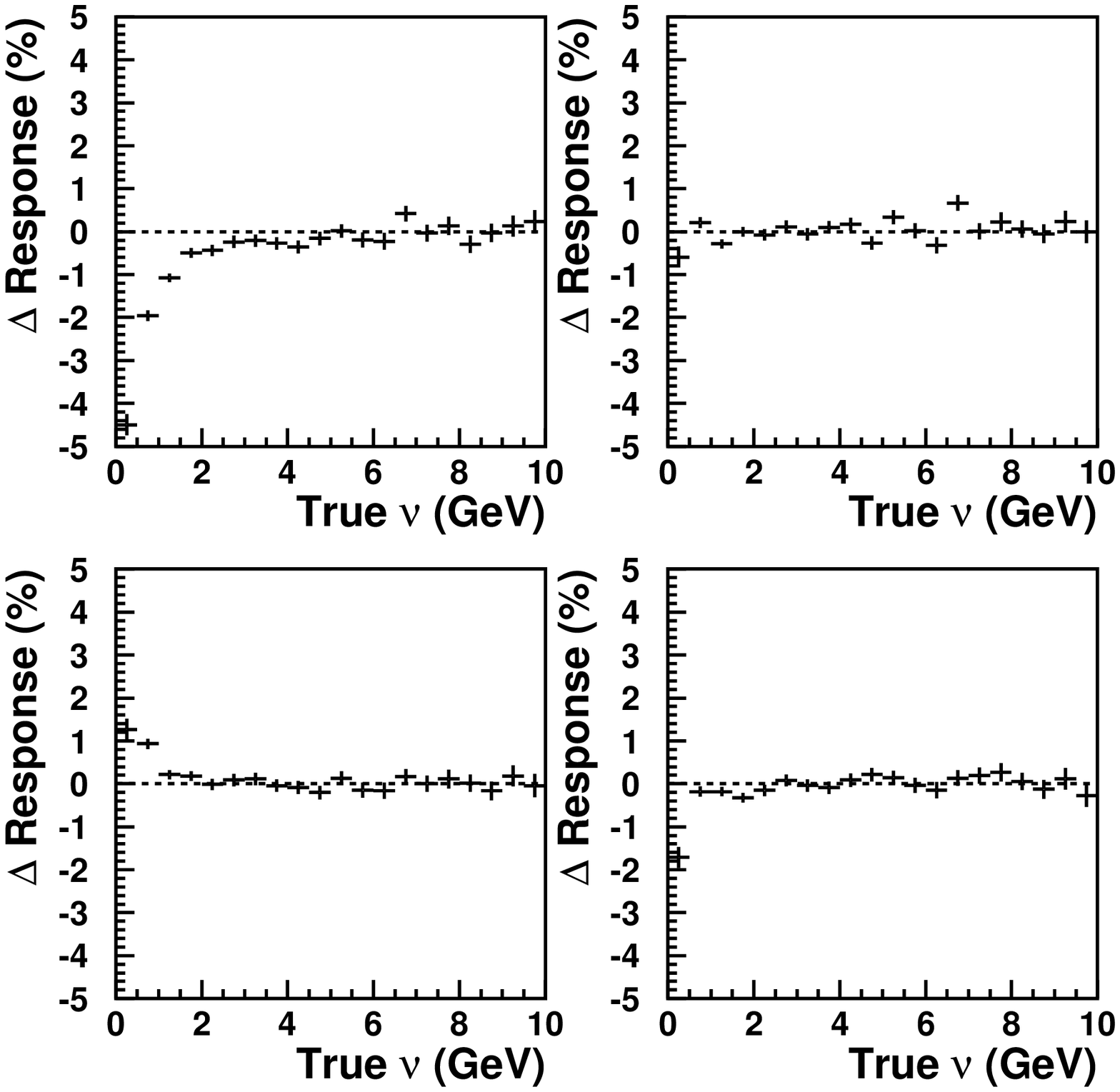}
  \caption{\label{fig:results8-11}
The results of $+1\sigma$ shifts to nucleon absorption (top left), secondary pion production 
by nucleons (top right), nucleon elastic scattering (bottom left), and the nucleon scattering
cross section (bottom right).}
\end{figure}

The change in the estimated shower energy in MINOS due to $+1\sigma$ changes in each of these inputs is shown in 
Figures \ref{fig:results2-5}, \ref{fig:results6-7}, and \ref{fig:results8-11}.  These results are in good agreement 
with those obtained using a different technique involving event-by-event reweighting of fully reconstructed MC events.   
The largest contribution coming from pion scattering is again the absorption process.  
The contributions from 
baryon rescattering are also sizable in the first few energy bins.  
For true shower energies of 0-500 MeV, the dominant process is quasi-elastic scattering.  
An increase to nucleon absorption produces a 4.5\% decrease in the response in this energy bin.  
Similarly an increase in nucleon elastic
scattering results in an increase in the response.  This is due to the fact that although an increase in the elastic
cross section is correlated with an increase in the total cross section, 71.9\% of nucleons already reinteract, so the 
increase in the total cross section has a relatively small effect on the number of interacted nucleons.  The elastic
increase is therefore largely offset by a decrease in the fraction of nucleons which are absorbed, which accounts for 
the increase in response.       

\begin{figure}
  \includegraphics[width=\columnwidth]{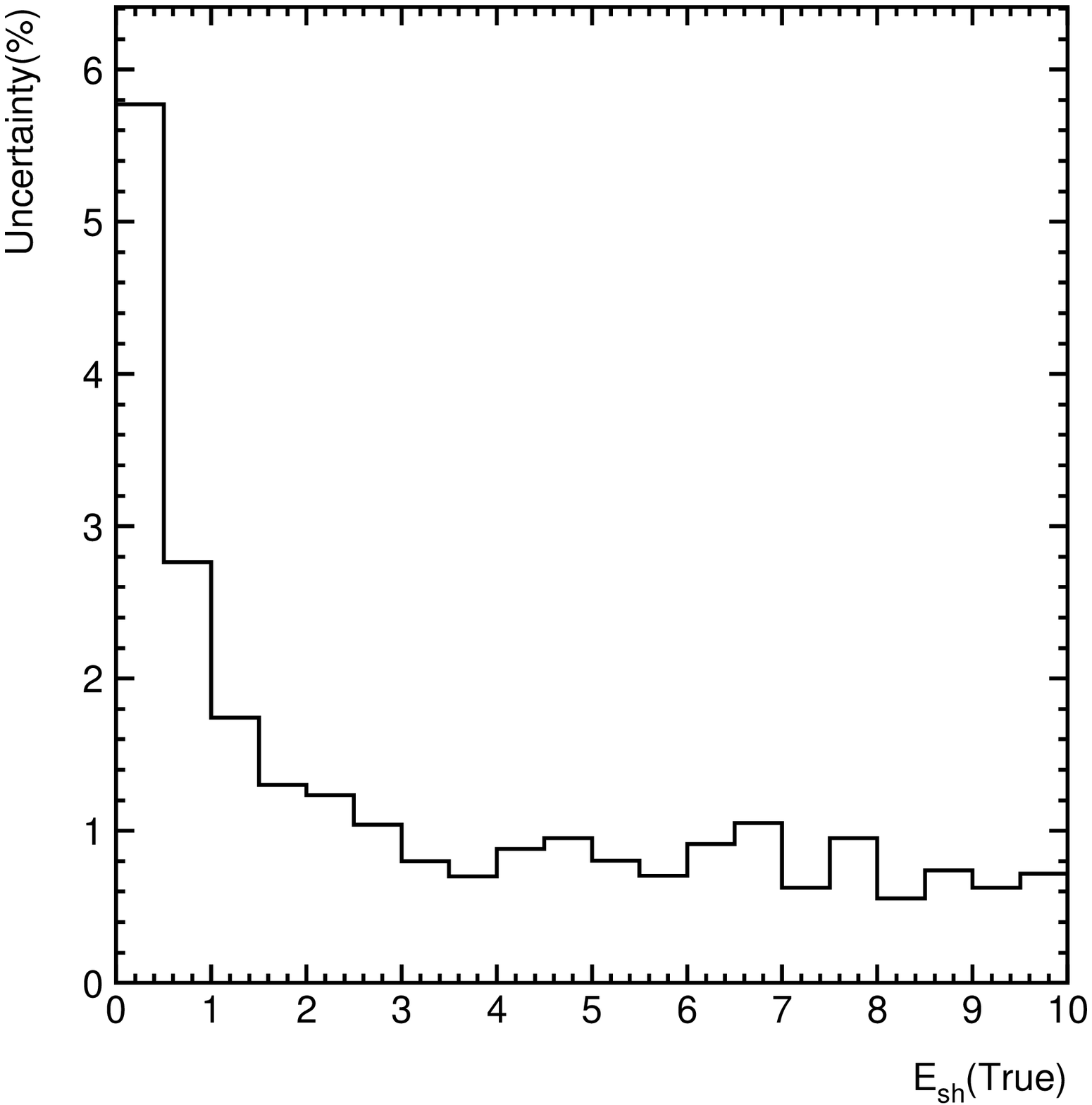}
  \caption{\label{fig:i1} Total uncertainty from all \inuke{} external data sources.}
\end{figure}

Figure~\ref{fig:i1} shows the contributions from all \inuke{} external data sources added in quadrature. 

\subsection{Model Assumptions}

In this work we also attempt to explicitly identify key assumptions in the models and 
evaluate their impact.  This task can be difficult for two reasons.  Firstly, assumptions in a model can be 
difficult to identify and may be all but invisible to any but the model creator.  Secondly it is not always
clear how to `undo' an assumption.  The best one can usually hope for is that there are two models, which 
incorporate {\bf different} assumptions, which can be compared side-by-side, and that there is abundant 
external data available to constrain the models.  

The first key assumption in \inuke{} is the way in which the free particle hadron scattering cross sections
are modified to account for the increased scattering cross section at low hadron energy.  
This change constituted the largest change to \inuke{} for \ngen{} \daikonplus.   The relevant \ngen{} parameter, EFNUCR
was tuned to its default value of 0.5 based on a comparison to pion-iron scattering data.  This setting was then
checked for consistency with neutrino data.  

There are numerous effects which one would expect to lead to different overall rescattering rates in pion and 
neutrino-induced reactions.  The most fundamental is that at low energies the deBroglie wavelength of the hadrons
is large and overlaps a significant fraction of the nucleus.  One would therefore expect differences in the 
scattering of pions approaching from infinity and pions born in the nuclear environment.   The external data 
(hadron and neutrino) is not sufficient to disentangle these differences since the statistics on the 
neutrino data are poor.   There is therefore a question of which external data one should use to determine the 
uncertainty on the EFNUCR parameter.  If one uses hadron data, the acceptable variation is $\pm0.10$, while 
if one uses neutrino data the allowable range is $\pm0.60$.  For this estimate we have taken the most conservative
approach and used the $0.60$ value.  

\begin{figure}
  \includegraphics[width=\columnwidth]{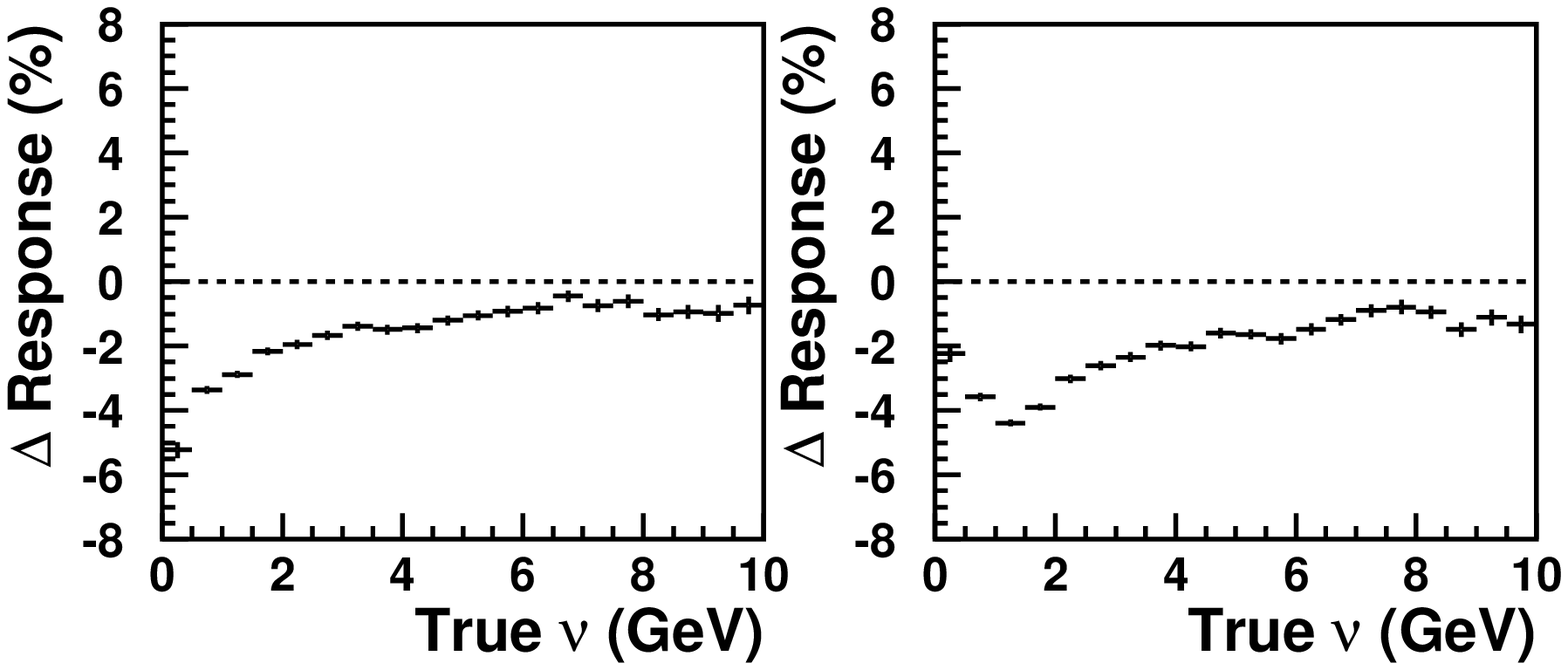}
  \caption{\label{fig:results12-13}
The results of a change to EFNUCR (left) and 
the treatment of pion/nucleon absorption (right).  Note scale change from previous plots.}
\end{figure}

The left hand panel of Figure \ref{fig:results12-13} shows the effect of changing EFNUCR from 0.50 to 1.10.  
For this comparison two other changes were also made to \ngen{}.  The first is a technical point - the increase in the 
nuclear size for low energy hadrons is normally limited to 0.75 times the nominal nuclear radius.  For these large 
changes to the model this cutoff value is removed.  The second is that in comparing with the neutrino data it is 
necessary to increase the pion absorption cross section to reach agreement for any value of EFNUCR.  The $\pm0.60$
allowable change to EFNUCR was determined including a $+0.5\sigma$ increase in the pion absorption cross section. 
Both of these effects are included in the simulated data set used for this calculation. 

The second major assumption has to do with the amount of missing energy produced in pion / nucleon absorption 
reactions.  When an absorption reaction occurs the pion energy is typically redistributed to 2-4 nucleons which 
carry the total energy of the absorbed particles.
These 'cascade' particles typically emerge from the nucleus followed by lower energy evaporation nucleons and 
de-excitation photons as the nucleus fragments or returns to the ground state.  \inuke{} only simulates the cascade
process and gives these nucleons the full pion energy in a phase space decay.  The dominant mechanism for missing
energy in this model is the production of low momentum nucleons which do not register in the detector.  

We have evaluated this assumption by making a dramatic and unphysical change to the model.  We produced a sample 
whereby pion and nucleon absorption produce eight-nucleon final states rather than the known four-nucleon state.  
The effect of this change is shown in the right panel of Figure \ref{fig:results12-13}.  

\begin{figure}
  \includegraphics[width=\columnwidth]{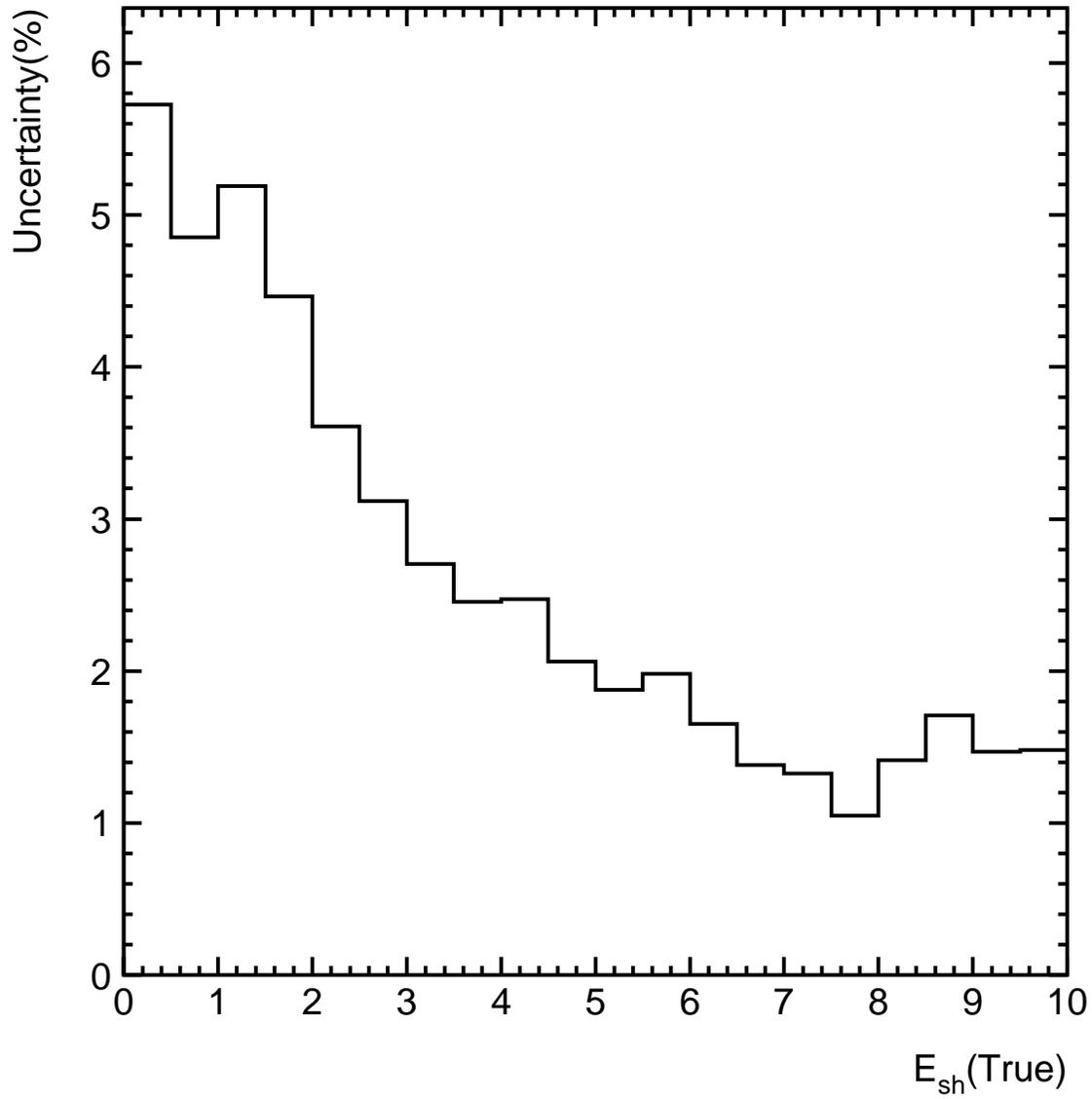}
  \caption{\label{fig:i2} Total uncertainty from all \inuke{} model assumptions.  }
\end{figure}

As Figure \ref{fig:results12-13} indicates, the model assumptions in \inuke{} have a large role in determining 
the overall energy scale.  Their quadrature sum is shown in Figure \ref{fig:i2}. 

\section{Formation Zone}

The treatment of the hadron formation zone used in \ngen{} is a model presented by the 
SKAT collaboration to describe their data \cite{Baranov:1984rv}.  
Previous studies which look at detailed measurements of hadron attenuation in electron
scattering experiments at the Jefferson Lab and 
HERMES have shown that this model fails to describe some important features, in particular the increase in nuclear
attenuation at high $z=E_\pi/\nu$.  

\begin{figure}
  \includegraphics[width=\columnwidth]{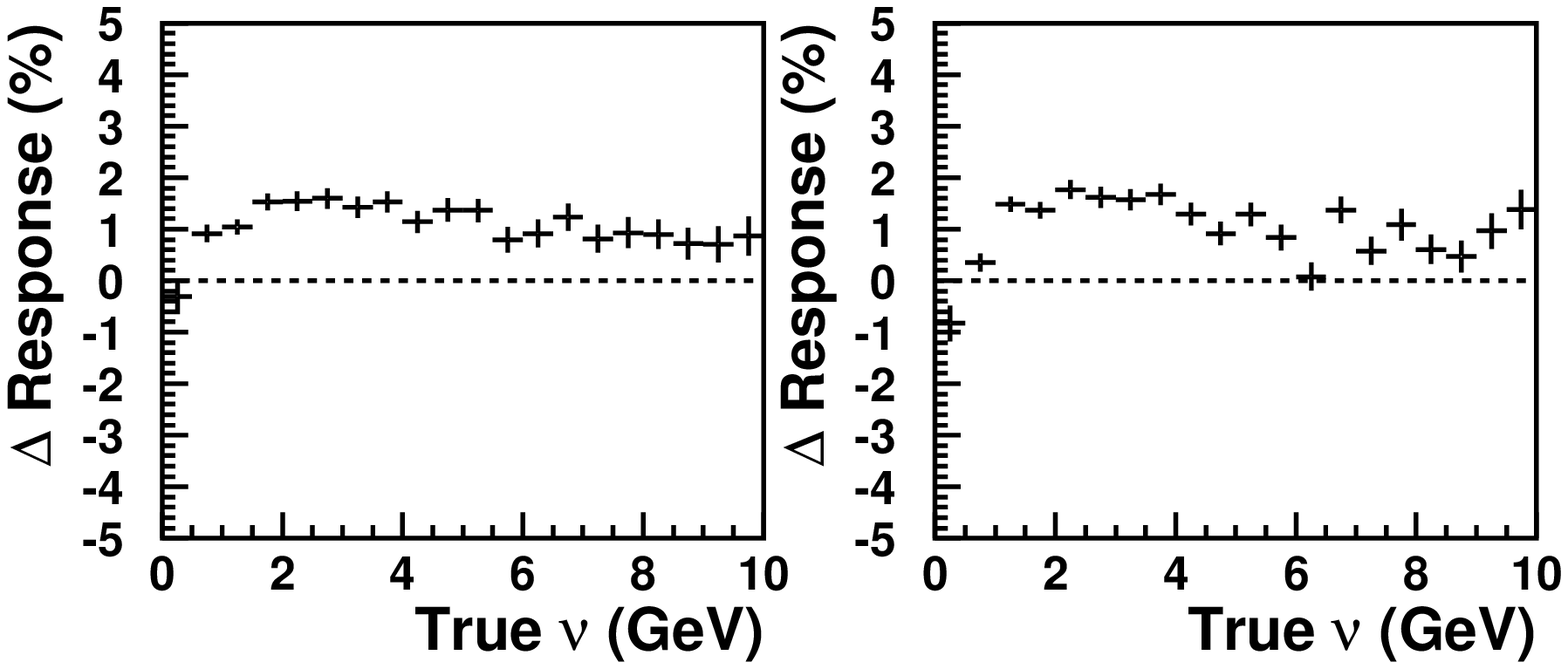}
  \caption{\label{fig:results14-15}
The results of a $+1\sigma$ change to the formation time (left) and 
use of a more sophisticated formation zone model (right).}
\end{figure}

Our estimates for formation zone uncertainty thus come from two pieces.  The first
is the measurement uncertainty on the single parameter in our current model, the formation time, as measured by the 
SKAT experiment \cite{Baranov:1984rv}.  This uncertainty is taken to be 50\%.  The effect of a +50\% increase in the 
formation time is shown in the left panel of Figure \ref{fig:results14-15}.  

The second contribution, the overall model uncertainty, has been evaluated by using a preliminary new model which 
is in better agreement with more recent data on hadron attenuation.  This model incorporates a more sophisticated 
modeling of the time development of the interaction cross section for hadrons produced in nuclei.  
This model is currently being compared with Jefferson Lab and HERMES data in preparation for 
inclusion in the next round of generator improvements. 
The change produced in going to the new model is shown in the right panel of Figure \ref{fig:results14-15}.   

\begin{figure}
  \includegraphics[width=\columnwidth]{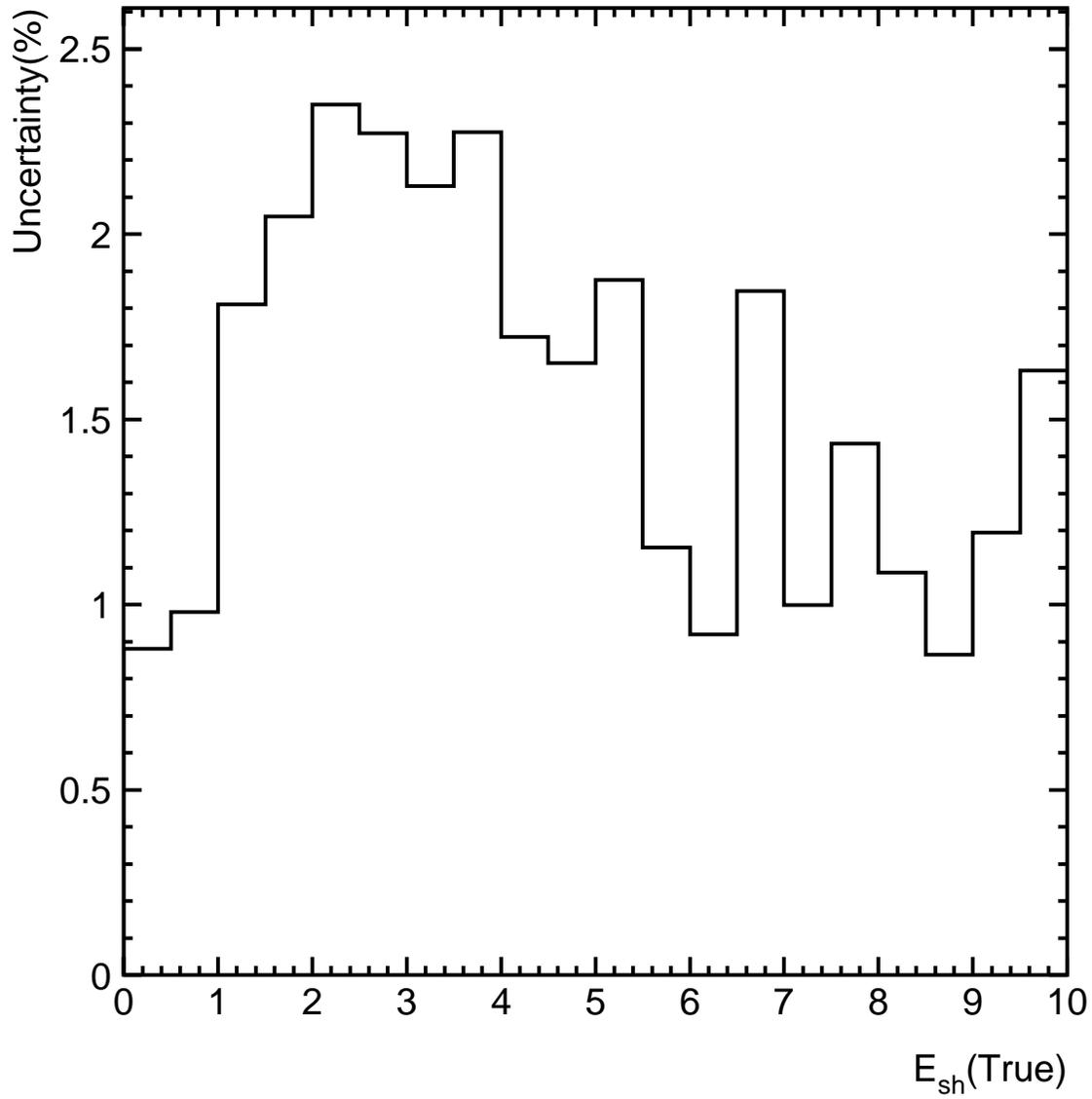}
  \caption{\label{fig:fz_total} Total uncertainty from all formation zone sources. }
\end{figure}

The quadrature
sum of the two contributions to the formation zone uncertainty is shown in Figure \ref{fig:fz_total}.  

\section{Hadronization Model Uncertainties}

\begin{figure}
  \includegraphics[width=\columnwidth]{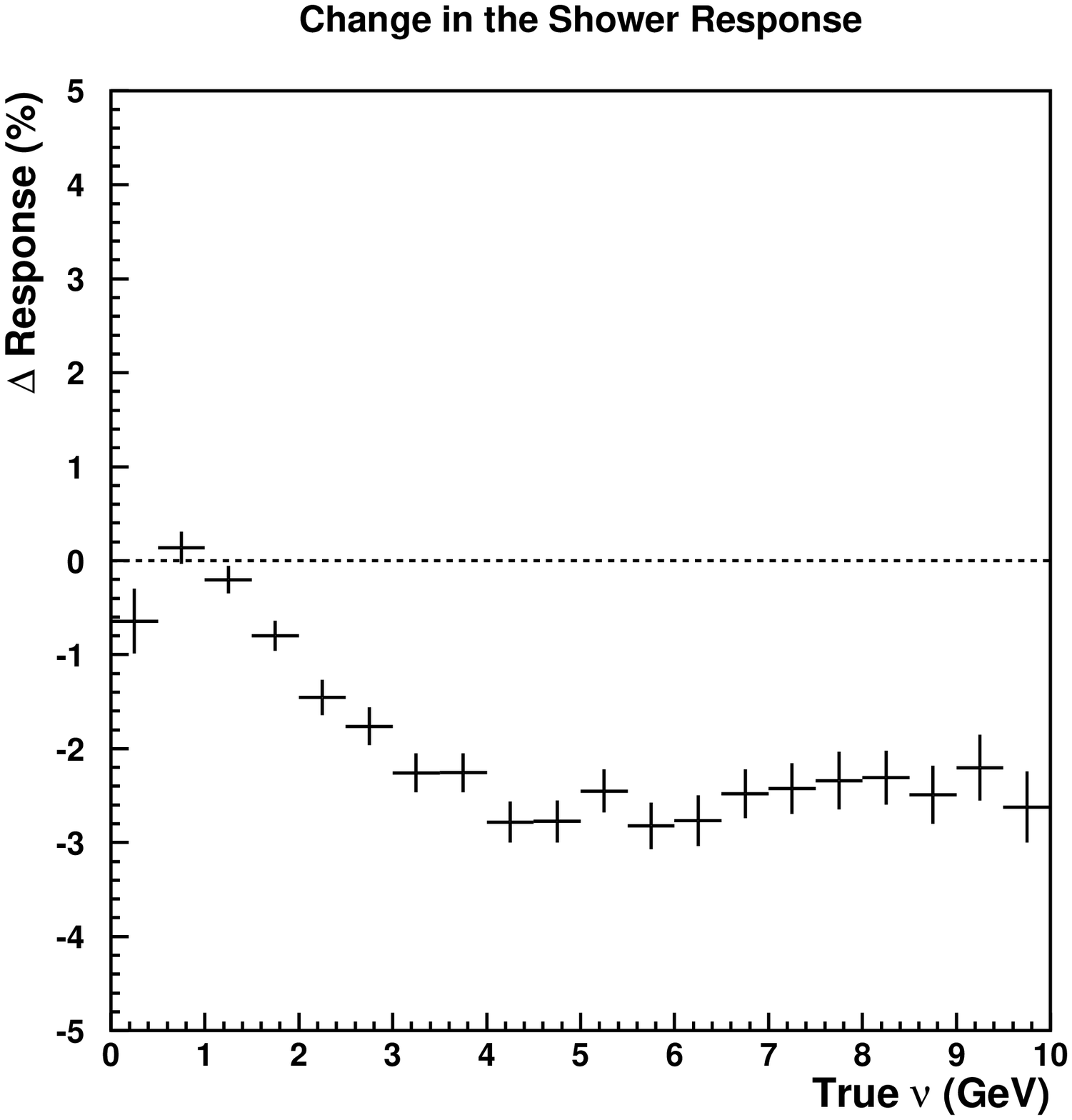}
  \caption{\label{fig:had}Changes to the hadronization model.}
\end{figure}

We evaluate the overall hadronization model uncertainty based on the change in going from the \carrot{} to 
\daikonplus{} hadronization model.    In this change many aspects of the model were changed which impact shower
energy scale.   These include particle multiplicities, the use of JETSET for hadronization, and the dynamics of the 
fragmentation 
process.  The \daikon{} version of \ngen{} does a much better job of describing the external data than the models in \carrot{} so this
change is conservative and certainly brackets any difference between the model and reality.  
For the evaluation of this component the evaluation sample was generated using the old hadronization model, but the 
rest of the simulation, in particular the intranuclear rescattering model, was the same as in \daikonplus{}. 
The uncertainty in the shower energy response coming from the hadronization model is shown in Figure \ref{fig:had}. 

\section{Conclusion}

\begin{figure}
  \includegraphics[width=\columnwidth]{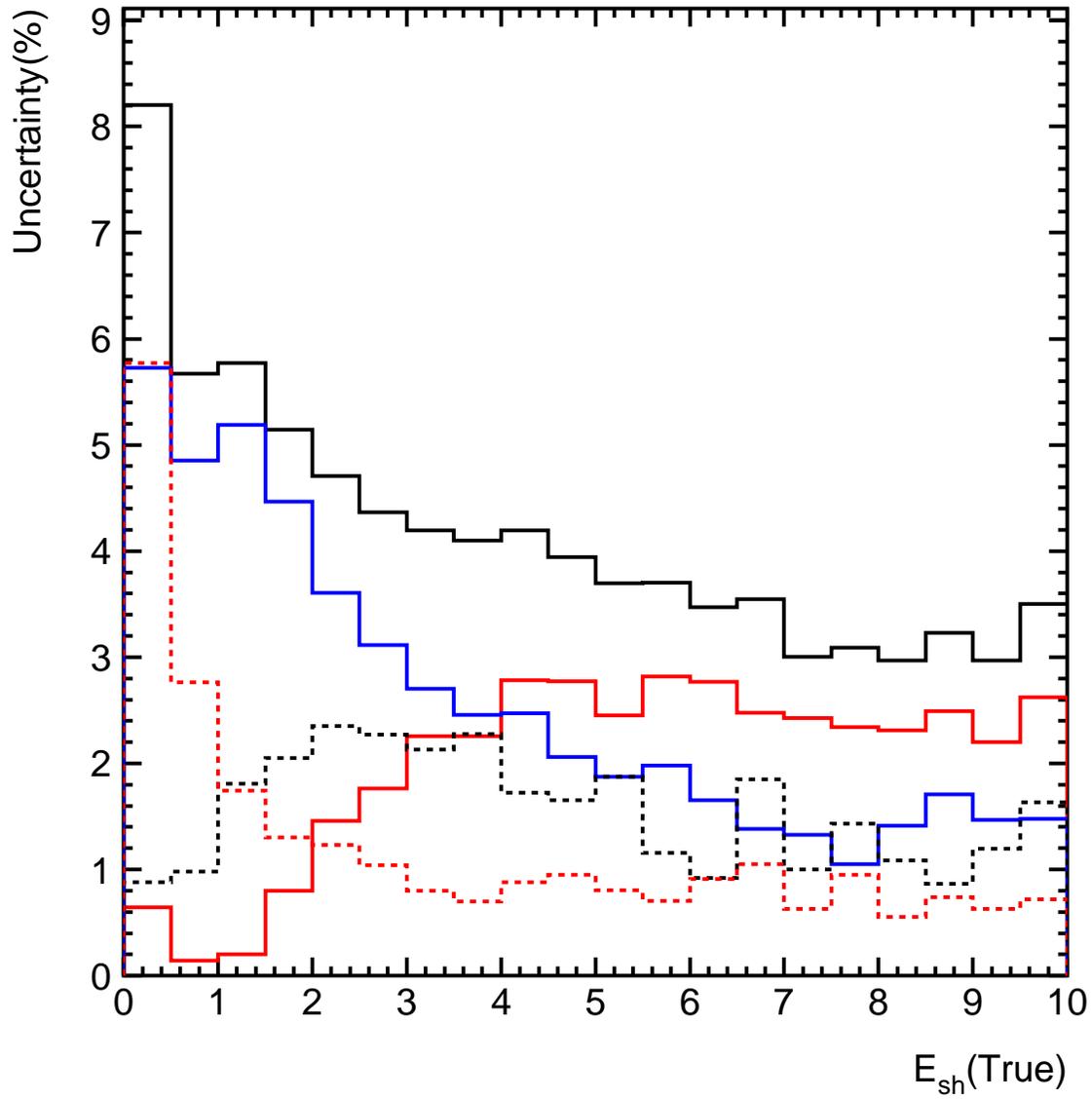}
  \caption{\label{fig:summary} Total uncertainty from all sources (solid black).  Contributions from intranuke assumptions (blue), 
\inuke{} input (dashed red), hadronization model (solid red), and formation zone (dashed black). }
\end{figure}

The quadrature sum of the contributions to the shower energy scale uncertainty from fifteen independent sources
is shown in Figure \ref{fig:summary}.  Also shown are the contributions from each of the categories considered in this
paper:  hadronization model, \inuke{} input data, \inuke{} assumptions, and formation zone.  The largest excursion
in a single bin is 8.2\% and occurs in the lowest energy bin.    

There is a strong energy dependence to the uncertainty.  The main reason is that the first two energy bins largely populated
by quasi-elastic events which are strongly affected by intranuclear rescattering for two reasons - the hadron energies are low and these
events are not subjected to the formation zone.  At high energies the uncertainty is reduced because the formation zone carries most 
of the hadrons out of the nucleus before they have a chance to interact.    For many MINOS analyses 
the hadronic energy scale uncertainty is characterized by a single number.  When that is done, a conservative approach is 
advocated where the 8.2\% value corresponding to the largest excursion a single energy bin should be used.  
\bibliography{hadsyst}
\bibliographystyle{unsrt}

\end{document}